\documentclass[conference]{IEEEtran}
\IEEEoverridecommandlockouts
\usepackage{cite}
\usepackage{amsmath,amssymb,amsfonts}
\usepackage{algorithmic}
\usepackage{graphicx}
\usepackage{textcomp}
\usepackage{xcolor}
\def\BibTeX{{\rm B\kern-.05em{\sc i\kern-.025em b}\kern-.08em
    T\kern-.1667em\lower.7ex\hbox{E}\kern-.125emX}}
    
\usepackage{makecell}
\usepackage[utf8]{inputenc}
\usepackage{graphicx}
\usepackage{algorithm}
\usepackage{algorithmic}
\usepackage{booktabs}
\usepackage{multirow}
\usepackage{xcolor}
\usepackage{hyperref}
\usepackage{tikz}
\usetikzlibrary{shapes,arrows,positioning}
\usepackage{listings}

\lstdefinestyle{prompt}{
  basicstyle=\ttfamily\scriptsize,
  frame=single,
  backgroundcolor=\color{gray!5},
  rulecolor=\color{black!20},
  breaklines=true,
  columns=fullflexible,
  keepspaces=true,
  showstringspaces=false,
  numbers=none,
  captionpos=t
  linewidth=0.95\linewidth,
  xleftmargin=0.02\linewidth,
  xrightmargin=0.02\linewidth,
}
\usepackage{caption}
\usepackage{float} 

\begin{document}

\title{Multi-Disciplinary Dataset Discovery from Citation-Verified Literature Contexts
\thanks{Zhiyin Tan was funded by the ``HybrInt - Hybrid Intelligence through Interpretable AI in Machine Perception and Interaction'' project (Zukunft Nds, Niedersächsisches Ministerium für Wissenschaft, Grant ID: ZN4219).
Changxu Duan was funded by the InsightsNet project (funded by the Federal Ministry of Education and Research (BMBF) under grant no. 01UG2130A). We gratefully acknowledge support from the hessian.AI Service Center (funded by the Federal Ministry of Research, Technology and Space, BMFTR, grant no. 16IS22091) and the hessian.AI Innovation Lab (funded by the Hessian Ministry for Digital Strategy and Innovation, grant no. S-DIW04/0013/003).
We gratefully acknowledge the computing time granted by the Resource Allocation Board and provided on the supercomputer Emmy/Grete at NHR-Nord@Göttingen as part of the NHR infrastructure. The calculations for this research were conducted with computing resources under the project nhr\_he\_starter\_25563. 
We acknowledge support from the SIGIR Student Travel Grant.}
}

\author{
\IEEEauthorblockN{Zhiyin Tan}
\IEEEauthorblockA{\textit{L3S Research Center}}
\textit{Leibniz University Hannover}\\
Hannover, Germany \\
zhiyin.tan@l3s.de \\
\and
\IEEEauthorblockN{Changxu Duan}
\textit{Technische Universität Darmstadt}\\
Darmstadt, Germany \\
duan@linglit.tu-darmstadt.de \\
}

\maketitle

\begin{abstract}
Identifying suitable datasets for a research question remains challenging because existing dataset search engines rely heavily on metadata quality and keyword overlap, which often fail to capture the semantic intent of scientific investigation. We introduce a literature-driven framework that discovers datasets from citation contexts in scientific papers, enabling retrieval grounded in actual research use rather than metadata availability. Our approach combines large-scale citation-context extraction, schema-guided dataset recognition with Large Language Models, and provenance-preserving entity resolution. We evaluate the system on eight survey-derived computer science queries and find that it achieves substantially higher recall than Google Dataset Search and DataCite Commons, with normalized recall ranging from an average of 47.47\% to a highest value of 81.82\%. Beyond recovering gold-standard datasets, the method also surfaces additional datasets not documented in the surveys. Expert assessments across five top-level Fields of Science indicate that a substantial portion of the additional datasets are considered high utility, and some are regarded as novel for the specific topics chosen by the experts. These findings establish citation-context mining as an effective and generalizable paradigm for dataset discovery, particularly in settings where datasets lack sufficient or reliable metadata. To support reproducibility and future extensions, we release our code, evaluation datasets, and results on GitHub \url{https://github.com/Fireblossom/citation-context-dataset-discovery}.
\end{abstract}

\begin{IEEEkeywords}
Dataset Discovery, Citation Context Mining, Scientific Literature Analysis, Cross-Disciplinary Research, Entity Resolution, Open Science
\end{IEEEkeywords}

\section{Introduction}

\begin{figure}
    \centering
    \includegraphics[width=0.80\linewidth]{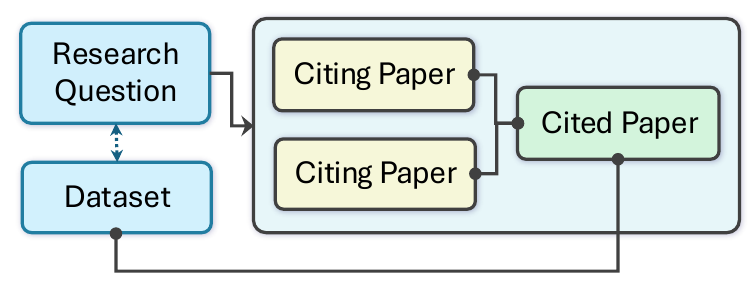}
    \caption{Bridging research questions to datasets via citation contexts.}
    \label{fig:pipeline_simple}
    \vspace{-5mm}
\end{figure}

Researchers across disciplines increasingly rely on publicly available datasets to drive knowledge discovery, ensure reproducibility, and facilitate subsequent data integration and reuse~\cite{wilkinson2016fair}. The number of open datasets continues to grow rapidly. DataCite~\cite{Brase2010} alone indexes over 30 million registered research objects, while Google Dataset Search~\cite{brickley2019google} covers millions more. 
 
Yet discovering suitable datasets for a given research problem remains difficult. Existing dataset search engines are largely metadata-driven.
Google Dataset Search matches queries against schema.org markup, while DataCite retrieves results from structured metadata fields. 
These systems work well when researchers know exact dataset names or precise terminology, 
But they struggle when metadata is incomplete, inconsistent, or too generic to capture nuanced applicability. Prior studies confirm that inconsistent metadata standards across repositories lead to fragmented discovery and unpredictable search outcomes~\cite{Zimmerman2007}.

In practice, researchers rarely begin with precise dataset names. 
Instead, they formulate research questions, often drawn from survey topics, prior studies, or emerging interdisciplinary problems, and seek datasets that can operationalize those questions. 
For such tasks, metadata keyword matching is insufficient. 
What is missing is contextual information: why a dataset was chosen, how it was used, and what research questions it helped answer.

We address this gap with a literature-driven dataset discovery framework. 
Our key insight is that scientific papers already contain rich contextual information about dataset usage. 
Citation contexts describe not only which dataset was used, but also how it supported the research design and why it was chosen. By treating these contexts as semantic bridges between research questions and datasets (see Figure~\ref{fig:pipeline_simple}), we can move beyond static metadata and ground dataset discovery in real-world research practice. This perspective naturally aligns with how researchers themselves work: when seeking data, they first look for papers similar to their goals, and then examine which datasets those papers employed.

Methodologically, we adapt techniques from scientific information extraction, originally developed for fine-grained entity recognition in scientific literature~\cite{luan-etal-2018-multi, jain-etal-2020-scirex}, and re-purpose them for multi-domain dataset discovery. This transfer allows us to circumvent the limitations of metadata dependency while anchoring recommendations in real-world usage patterns documented in scientific literature.

Our contributions are as follows:
\begin{itemize}
\item We introduce a literature-driven dataset discovery framework that bypasses metadata dependency by leveraging usage contexts documented in scientific papers.
\item We develop a three-stage computational pipeline that integrates neural language models with citation graph analysis for scalable multi-domain dataset extraction.
\item We establish a rigorous and scalable evaluation protocol covering all six top-level Fields of Science and Technology (FOS)\footnote{FOS: Natural Sciences, Engineering and Technology, Medical and Health Sciences, Agricultural Sciences, Social Sciences, and Humanities.}. For Computer Science, we perform an automated recall-based evaluation using gold-standard datasets derived from survey papers. To ensure generality, we evaluate the remaining five FOS domains through expert assessments of relevance, utility, and novelty.
\item We demonstrate that our approach substantially improves recall over leading dataset search engines, and that the additional datasets it surfaces are often judged by domain experts as high utility and, in many cases, novel for their research areas.
\end{itemize}

This work addresses a practical need, enabling researchers to efficiently identify relevant datasets in unfamiliar domains. At the same time, it contributes to the broader vision of intelligent scientific knowledge organization.

\section{Related Work}

Our work addresses dataset discovery through systematic analysis of scientific literature. We review three relevant areas: existing dataset search systems, literature-based information extraction, and citation context analysis.

\subsection{Dataset Search Engines and Repositories}

Existing dataset discovery systems include general-purpose search engines such as Google Dataset Search~\cite{brickley2019google}, which relies on schema.org markup, and DataCite~\cite{Brase2010}, which ensures metadata quality through DOI registration but indexes only formally published datasets. General-purpose repositories such as Zenodo, Kaggle, and Mendeley Data accept contributions across disciplines, while domain-specific repositories such as Protein Data Bank (PDB), Gene Expression Omnibus (GEO), Inter-university Consortium for Political and Social Research (ICPSR), and British National Corpus (BNC) provide curated, high-quality resources within particular communities but remain siloed, limiting interdisciplinary discovery.  

These systems share fundamental limitations: heavy reliance on keyword matching against metadata, lack of semantic understanding of research contexts, and terminological mismatches across disciplines~\cite{borgman2015big,pasquetto2017uses}. Few systems address connecting abstract research questions with datasets when metadata is sparse or inconsistent.

\subsection{Literature-Based Information Extraction}

Scientific papers capture contextual details about resource usage that metadata alone can not~\cite{chen2019survey}. Prior work has leveraged this signal to build dataset mention extraction benchmarks:
SciERC~\cite{luan-etal-2018-multi} focused on computer science papers, SciREX~\cite{jain-etal-2020-scirex} extended extraction to the document level, and DMDD~\cite{pan-etal-2023-dmdd} provided large-scale multi-domain coverage.

These efforts demonstrate the feasibility of mining dataset mentions at scale, but their primary goal is catalog construction or usage analysis rather than guiding researchers to datasets relevant to new questions~\cite{Heddes2021Detection}. Full-document extraction can also be computationally expensive, making real-time search and recommendation challenging \cite{xu2025ChatPD, marini-etal-2025-data, dme_jcdl2024}. 
These methods are also limited by their dependency on high-quality structured inputs and a lack of robust cross-disciplinary evaluation.

We build on this line of work by treating extracted mentions not as endpoints for cataloging but as semantic bridges linking research questions to datasets. In doing so, we leverage citation contexts to provide multi-domain applicability and enable efficient, literature-driven dataset discovery.

\subsection{Citation Context Analysis}

Citation context analysis reveals semantic relationships between resources and applications. Zhao et al.~\cite{zhao-etal-2019-context} developed frameworks for classifying citation functions, while Färber et al.~\cite{Farber2021UsedVsMentioned} distinguished between actively used versus mentioned datasets. Recent advances show neural language models excel at scientific information extraction~\cite{marini-etal-2025-data}, and frameworks like SOFT~\cite{DuanTan2025SOFT} disentangle citation intent from content type.

However, existing approaches treat context as auxiliary information rather than the primary semantic signal for discovery. We extend these techniques by treating citation contexts as semantic bridges connecting research questions to datasets across disciplines, creating a generalizable framework that operates without domain-specific training by leveraging scientific literature as a comprehensive knowledge base of dataset usage patterns~\cite{cohan-etal-2020-specter,wilkinson2016fair}.

\begin{figure*}
    \centering
    \includegraphics[width=0.99\linewidth]{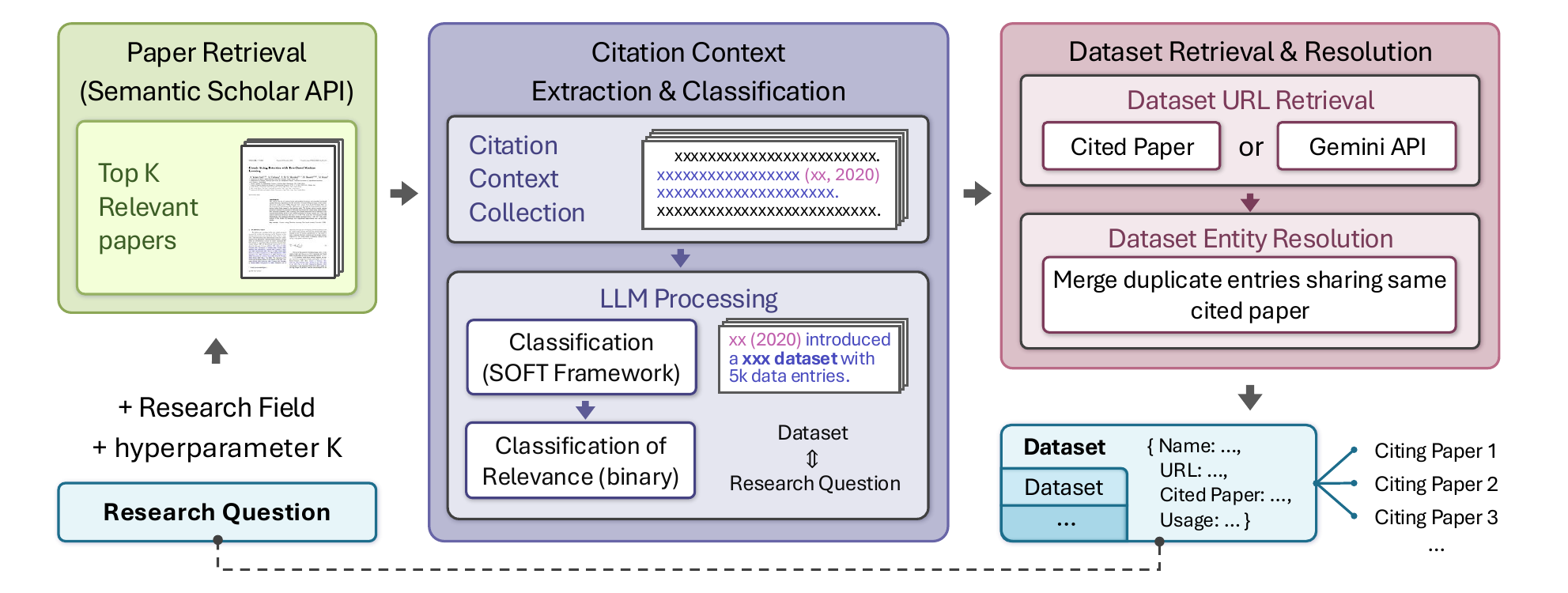}
    \caption{The proposed pipeline maps a research question to relevant datasets via citation context analysis and metadata linking.}
    \label{fig:pipeline}
\end{figure*}

\section{Methodology}

We formulate dataset discovery as \emph{contextual information extraction} over scientific literature. The core idea is to use citation contexts as semantic evidence linking natural-language research questions to datasets. Our system is a three-stage pipeline that combines efficient corpus access with neural extraction and robust entity resolution: 
(1) scalable citation-context retrieval, 
(2) neural dataset mention extraction with citation-aware quality signals, and 
(3) datasets entity resolution for consolidation. This design directly addresses three chronic shortcomings of metadata-only systems: \emph{semantic mismatch}, \emph{contextual ambiguity}, and \emph{incomplete coverage}.

\subsection{Problem Formulation}

Let a research query $\mathcal{Q}$ (free text) and optional field constraints $\mathcal{F}$ be given. The task is to retrieve and rank dataset entities $\mathcal{D}=\{d_1,\dots,d_n\}$ from a literature corpus.
In this work, we operationally define a dataset $d$ as any named research resource (e.g., corpora, benchmarks, databases) that is explicitly mentioned in a paper as being used, modified, or evaluated against. This definition excludes mentions of software tools or general methodologies.

Each entity $d_i$ is a structured tuple
\begin{align}
d_i=\langle \text{name}_i,\ \mathcal{C}_i,
\ \mathcal{M}_i\rangle,
\end{align}
where $\text{name}_i$ is the canonical identifier, $\mathcal{C}_i$ is the set of supporting citation contexts, 
and $\mathcal{M}_i$ contains enriched metadata.


\subsection{Three-Stage Pipeline}

\subsubsection{Stage 1: Scalable citation-context retrieval}
We operate on Semantic Scholar Academic Graph (S2AG)~\cite{kinney2025semanticscholaropendata} with a hybrid strategy: precomputed indices for citation windows plus on-demand filtering for query specialization. Given $(\mathcal{Q},\mathcal{F})$, we retrieve candidate papers, extract sentence-level windows around citation markers, and retain $(\text{paper},\text{target},\text{window})$ triples with minimal metadata needed downstream. To maintain throughput over millions of papers while keeping latency manageable, we combine (1) prebuilt context tables and (2) light query-time joins; we further reduce noise with an LLM-based relevance filter in a \texttt{retrieve-then-read} setup~\cite{marini-etal-2025-data}. This balances accuracy with computational cost (full-document processing is expensive for real-time recommendation). 

\subsubsection{Stage 2: LLM-based dataset mention extraction with citation-aware quality}
We cast extraction as Dataset Mention Extraction \cite{dme_jcdl2024}
over citation windows, augmented with citation-function cues. Building on scientific information retrieval~\cite{jain-etal-2020-scirex,zhang-etal-2024-scier}, we adapt Qwen2.5-72B-Instruct~\cite{qwen2.5}, a large instruction-tuned model, to identify dataset mentions, including abbreviated or implicit references. 
To output a structured record with (1) cited content type and (2) citation intent (following SOFT’s \cite{DuanTan2025SOFT} separation of intent vs. content type). This structured extraction captures the specific application context of a dataset, such as whether it was primarily \textit{used} as a resource, \textit{modified} for a new task, or served as a standard to \textit{evaluate against}. Such semantic insight provides a level of granularity that is fundamentally unavailable to metadata-driven search systems and allows researchers to filter resources by their function in prior research.
We enforce a JSON schema for downstream consistency and apply a multi-tier validator (schema checks, semantic checks, domain rules) to the extraction quality. We also compute citation-enhanced signals. e.g., recency-weighted usage and context salience, to complement model confidence.
A high-level procedure is given in Algorithm~\ref{alg:neural_extraction}.

\begin{algorithm}[!t]
\caption{LLM-based Dataset Mention Extraction}
\label{alg:neural_extraction}
\begin{algorithmic}[1]
\REQUIRE Citation windows $\mathcal{C}$, query/topic hints $\mathcal{T}$
\ENSURE Structured extractions $\mathcal{E}$ with confidence and evidence
\STATE $\mathcal{E} \leftarrow \varnothing$
\FOR{$c \in \mathcal{C}$}
  \STATE $y \leftarrow \text{LLMExtract}(c,\mathcal{T})$ \COMMENT{datasets, intent, content type}
  \STATE $z \leftarrow \text{Validate}(y)$ \COMMENT{schema/semantic/domain}
  \STATE $r \leftarrow \text{RelevanceFilter}(z,\mathcal{T})$ \COMMENT{second LLM pass}
  \IF{$r$ is valid}
    \STATE $\mathcal{E} \leftarrow \mathcal{E} \cup \{\text{Record}(r.\text{items}, r.\text{confidence}, r.\text{evidence})\}$
  \ENDIF
\ENDFOR
\RETURN $\mathcal{E}$
\end{algorithmic}
\end{algorithm}

\subsubsection{Stage 3: Dataset entity resolution and integration}
We adopt a deterministic, evidence-preserving consolidation pipeline rather than graph clustering. First, mentions tied to the same bibliographic relation are locally consolidated by selecting the most reliable mention and unioning compatible attributes (e.g., roles, brief descriptions), while recording the number of merged mentions for diagnostics. Second, we perform a global aggregation by a strict canonical name: surface forms are normalized by trimming quotation/bracket marks, removing parenthetical qualifiers and generic type words (e.g., dataset/corpus/benchmark or train/test/validation markers), collapsing whitespace, stripping punctuation, and lowercasing. Mentions sharing the same canonical name are grouped; we retain all provenance, select a human-readable display name by frequency, and keep all surface variants as aliases. References are rendered with persistent identifiers preferred over URLs.
The procedure is summarized in Algorithm~\ref{alg:entity_resolution}.

\begin{algorithm}[!t]
\caption{Deterministic Entity Consolidation}
\label{alg:entity_resolution}
\begin{algorithmic}[1]
\REQUIRE Mentions $\mathcal{E}$ (surface name, relation, attributes, provenance)
\ENSURE Consolidated entities $\hat{\mathcal{E}}$
\STATE $\mathcal{E}' \leftarrow \text{LocalConsolidateByRelation}(\mathcal{E})$
\FOR{$m \in \mathcal{E}'$} \STATE $m.\text{key} \leftarrow \text{Normalize}(m.\text{name})$ \ENDFOR
\STATE $G \leftarrow \text{GroupByKey}(\mathcal{E}')$
\FOR{$g \in G$}
  \STATE $display \leftarrow \text{SelectDisplayName}(g.\text{names})$
  \STATE $aliases \leftarrow \text{Unique}(g.\text{names})$
  \STATE $evidence \leftarrow \text{CollectProvenance}(g)$
  \STATE $links \leftarrow \text{PreferPID}(\text{CollectLinks}(g))$
  \STATE $\hat{\mathcal{E}} \leftarrow \hat{\mathcal{E}} \cup \{\text{Entity}(display, aliases, evidence, links)\}$
\ENDFOR
\RETURN $\hat{\mathcal{E}}$
\end{algorithmic}
\end{algorithm}

Putting the stages together yields an end-to-end flow from query to ranked datasets.
This architecture scales by design (preindexed contexts and light query-time filtering), preserves accuracy (LLM-based extraction and citation-aware signals), and delivers reliable outputs, resolution, and provenance.

\subsection{Evaluation Framework and Quality Metrics}
\label{sec:evaluation}

We evaluate along \emph{extraction quality}, \emph{evidence quality}, and \emph{practical accessibility}, following FAIR~\cite{wilkinson2016fair}, Information Retrieval evaluation~\cite{Manning2008}, and data quality standards~\cite{batini2009methodologies}. Automated metrics are computed against survey-derived gold standards in computer science, while expert assessments are used to judge the utility and novelty of additional datasets that fall outside these gold-standard sets in other FOS domains.

\subsubsection{Extraction quality}
Our evaluation employs a multi-perspective approach with three matching granularities, inspired by entity resolution evaluation practices~\cite{christen2012data}. It measures both extraction effectiveness and evidence quality, following scientific data evaluation standards~\cite{koesten2017dataset}.
\begin{itemize}
\item \textit{Exact Recall}: exact string match after whitespace normalization.
\item \textit{Norm Recall}: canonical normalization (lowercase, punctuation removal, whitespace consolidation).
\item \textit{Fuzzy Recall}: sequence-similarity clustering with threshold $\tau=0.9$~\cite{cohen2003comparison}.
\end{itemize}

The primary evaluation metric is \textit{Norm Recall}, which provides the optimal balance between precision and robustness for practical applications, following best practices in scientific dataset evaluation~\cite{koesten2017dataset}. 
We also report 
\begin{equation}
    \textbf{FuzzyGain} = \textit{Fuzzy Recall} -\textit{Norm Recall}
\end{equation} to quantify robustness to name variation.

\subsubsection{Evidence quality}
\begin{itemize}
  \item \textit{Trusted Sources (\%)}: share of matched entities whose evidence resolves to authoritative provenance (PID landing pages or trusted hosts such as official catalogs/registries or maintainer repositories).
  \item \textit{With DOI (\%)}: share of entities with a persistent identifier (DOI/HANDLE/ARK); reported as “With DOI (\%)” in our tables.
\end{itemize}

\subsubsection{Practical accessibility and efficiency}
We measure Redundancy to evaluate the effectiveness of entity consolidation, where a lower score indicates successful merging of duplicate mentions referring to the same dataset:
\begin{equation}
    \text{Redundancy}=\frac{|\text{Mentions}|-|\text{Entities Norm}|}{\max(1,|\text{Entities Norm}|)}
\end{equation}
Gold standards are expert-curated, domain-specific reference sets with standardized protocols for consistency across areas; we compare system recall directly against these ground truths.

\subsection{Reproducibility}
All code, configs, and evaluation workbooks are released (Section~\ref{sec:experiments}); implementation particulars that affect replication are documented in Appendices~\ref{app:stage1}–\ref{app:stage3}.

\subsection{Pipeline Workflow Visualization}

Figure~\ref{fig:pipeline} illustrates the complete workflow of our three-stage dataset discovery pipeline. The system begins with a research query and field constraints as input, progresses through citation context extraction, neural dataset extraction with citation-enhanced quality assessment, and entity resolution, ultimately producing a ranked dataset table. The optional LLM pre-filtering stage enables domain-specific relevance assessment to reduce noise in large-scale retrieval scenarios.

\section{Experiments}
\label{sec:experiments}

\subsection{Experimental Setup}

We evaluate our literature-driven dataset discovery framework under two complementary settings. The first is an automated, large-scale benchmark in computer science, where survey papers provide natural entry points for dataset queries. The second is a cross-disciplinary expert study, designed to test the practical utility of our approach beyond computer science. Together, these settings answer three research questions: 
(1) How does our pipeline compare to existing approaches in coverage and quality? 
(2) What is the practical utility of retrieved datasets as judged by domain experts? 
(3) How do individual components contribute to performance?

\subsubsection{Survey Paper–Derived Queries (computer science Benchmarks)}
For large-scale automated evaluation, we focused on computer science domains where curated surveys reflect active research questions. We selected eight representative survey papers from top-tier venues in natural language processing (ACL, EMNLP, NAACL) and computer vision (CVPR, ICCV, TPAMI). Each survey title was used as a natural-language query, ensuring that our evaluation queries represent authentic, community-defined research problems rather than artificially constructed prompts. This process yielded the eight benchmark tasks detailed alongside their results in Table~\ref{tab:domain_results_revised}. The gold standard for each task was manually extracted from the corresponding survey paper to serve as our ground truth for the automated evaluation.

\subsubsection{Expert-Provided Queries (Cross-Disciplinary Evaluation)}

To validate cross-domain applicability, we further evaluate on research questions sourced directly from experts in diverse fields. Following FOS taxonomy~\cite{sinha2015overview}, we exclude computer science (our own domain) to avoid bias, and select research questions based on the remaining top-level categories: Engineering and Technology, Medical and Health Sciences, Agricultural Sciences, and Social Sciences and Humanities. Domain experts (Ph.D. holders, doctoral candidates, and advanced master's students under doctoral supervision) were asked to propose a question of genuine research interest within their discipline. These queries anchor our expert evaluation protocol (Section~\ref{sec:expert_eval}), ensuring that the assessment captures both the generality of our framework and its practical value in real-world, cross-disciplinary discovery.

\begin{table*}[!t]
\centering
\caption{Statistical-based performance evaluation across all computer science tasks. $\uparrow$ indicates a higher value for better performance, $\downarrow$ indicates a lower value for better performance.}
\label{tab:main_results}
\begin{tabular}{@{}p{3.cm}r>{\raggedleft\arraybackslash}p{1.8cm}r>{\raggedleft\arraybackslash}p{1.8cm}r>{\raggedleft\arraybackslash}p{1.75cm}@{}}
\toprule
\textbf{Method} & \textbf{\makecell[r]{Total\\ Entities}} & \textbf{\makecell[r]{FuzzyGain\\ (\%) $\uparrow$}} & 
\textbf{\makecell[r]{Trusted Sources \\ (\%) $\uparrow$}} &
\textbf{\makecell[r]{With DOI\\ (\%) $\uparrow$}} & 
\textbf{\makecell[r]{Redundancy \\ $\downarrow$}} & 
\textbf{\makecell[r]{Evidence \\ Quality}} \\
\midrule
\textbf{Ours}    & \textbf{1,330} & \textbf{1.71} & \textbf{14.23} & 68.52 & \textbf{0.26} & \textbf{High} \\
Google Dataset Search        & 79             & 0.00           & 1.63           & 32.81 & 0.29          & Medium \\
DataCite Commons         & 67             & 0.00           & 0.00           & \textbf{87.50} & 7.76          & \textbf{High} \\
\bottomrule
\end{tabular}
\end{table*}

\begin{table*}[!htbp]
\centering
\caption{For each of the eight computer-science queries, we report the number of gold-standard datasets (constructed from survey papers), the number of matched datasets, and the corresponding macro-averaged recall (\%).}
\label{tab:domain_results_revised}
\begin{tabular}{l r r r r r r r}
\toprule
\multirow{2}{*}{\textbf{Research Question}} 
& \multirow{2}{*}{\textbf{Gold}} 
& \multicolumn{3}{c}{\textbf{Matched}} 
& \multicolumn{3}{c}{\textbf{Recall (\%)}} \\
\cmidrule(lr){3-5} \cmidrule(lr){6-8}
& & \textbf{Ours} & \textbf{Google} & \textbf{DataCite}
  & \textbf{Ours} & \textbf{Google} & \textbf{DataCite} \\
\midrule
Multi-modal Knowledge Graph Reasoning~\cite{Liang2024} & 11 & 9 & 0 & 0 & 81.82 & 0.00 & 0.00 \\
All-in-One Image Restoration~\cite{Jiang2025} & 30 & 10 & 0 & 0 & 33.33 & 0.00 & 0.00 \\
Planning Capabilities of LLM~\cite{wei-etal-2025-plangenllms} & 38 & 21 & 0 & 0 & 55.26 & 0.00 & 0.00 \\
Event-based Stereo Depth Estimation~\cite{Ghosh2025} & 17 & 9 & 0 & 0 & 52.94 & 0.00 & 0.00 \\
Patent Classification in NLP~\cite{shomee-etal-2025-survey} & 7 & 3 & 0 & 0 & 42.86 & 0.00 & 0.00 \\
Document-level Event Extraction~\cite{zheng-etal-2024-comprehensive} & 23 & 9 & 3 & 0 & 39.13 & 13.04 & 0.00 \\
Text Line Segmentation for Historical Documents~\cite{Rabaev2025} & 43 & 13 & 1 & 0 & 30.23 & 2.33 & 0.00 \\
Personalized Text Generation~\cite{xu-etal-2025-personalized} & 16 & 1 & 1 & 0 & 6.25 & 6.25 & 0.00 \\
\midrule
\textbf{Average Recall} & & & & & \textbf{47.47} & 2.70 & 0.00 \\
\bottomrule
\end{tabular}
\end{table*}

\subsection{Baseline Systems}

\textbf{Google Dataset Search~\cite{brickley2019google}} is the most comprehensive web-scale dataset search engine, maintained by Google and indexing millions of datasets through schema.org markup. As the de facto standard for dataset discovery, it provides broad coverage across domains but relies heavily on the quality of metadata.

\textbf{DataCite Commons~\cite{Brase2010}} is the authoritative registry for research data DOIs, maintained by a global consortium of institutions and libraries. With over 30 million registered objects, it represents the gold standard for formally published datasets, though coverage is biased toward curated resources.

While baseline systems target metadata search, our model generalizes to dataset and application reasoning.

\subsection{Expert Evaluation Protocol}
\label{sec:expert_eval}

Automated metrics cannot capture contextual appropriateness or real-world utility~\cite{borgman2015big}. We therefore conduct a double-blind expert study~\cite{Manning2008}, based on expert-provided queries from non-CS domains.

\subsubsection{Participant Recruitment}
We invited ten domain experts: three Ph.D. holders, four doctoral candidates, and three advanced master's students under doctoral supervision. 
Disciplines include Engineering and technology (Food and beverages), Medical and Health Sciences (Clinical medicine), Agricultural sciences (Agriculture, Forestry, and Fisheries), Social Sciences (Educational sciences), and Humanities (Arts). The comparative performance of each system on these expert-provided queries is visualized in Figure~\ref{fig:rq_radars}, providing a qualitative counterpart to our automated benchmarks.

\subsubsection{Anonymization Strategy}

To prevent bias, results from the three systems are presented anonymously as ``System A, B, and C". Experts are unaware of the underlying methodologies.

\subsubsection{Evaluation Dimensions}

Experts rated candidate datasets, including both survey-documented and additional datasets surfaced by each system, on six 5-point scales:
    \begin{itemize}
        \item \textit{Relevance:} Alignment with the stated research question
        \item \textit{Utility:} Likelihood of practical usage in research
        \item \textit{Accessibility:} Clarity of descriptions and ease of access
        \item \textit{Trustworthiness:} Confidence in reliability and source quality
        \item \textit{Novelty:} Discovery of previously unknown datasets
        \item \textit{Overall Satisfaction:} Holistic system assessment
    \end{itemize}

\subsubsection{Data Collection and Analysis}
Responses were collected via anonymized Google Forms. Ratings were averaged within each expert–system pair and then aggregated across experts. Given the small and unbalanced number of raters per query, we did not conduct formal significance testing. We report per-dimension means and best‑vote preferences across experts, which complement automated benchmarks with domain‑grounded qualitative evidence. While sample sizes limit statistical power, results complement automated benchmarks by providing domain-grounded perspectives on dataset utility.

\subsection{Results and Analysis}

We present comprehensive experimental results comparing our neural dataset discovery approach against two established baseline systems: Google Dataset Search and DataCite Commons. Our evaluation encompasses eight diverse research domains, analyzing 1,358 total dataset mentions across multiple quality dimensions.

\subsubsection{Overall Automated Performance}

Table~\ref{tab:main_results} summarizes the aggregated automated evaluation for the computer science benchmarks. Our citation-context approach extracts 1,330 unique dataset entities and achieves substantially higher average normalized recall (47.47\%) than Google Dataset Search (2.70\%) and DataCite (0.00\%), while maintaining high evidence quality and low redundancy. Table~\ref{tab:domain_results_revised} reports per-query results, with recall for our system reaching up to 81.82\% on individual survey-derived tasks.

\subsubsection{Expert Evaluation Results}

\begin{figure*}[!t]
\vspace{-5mm}
\centering
\includegraphics[width=0.9\linewidth]{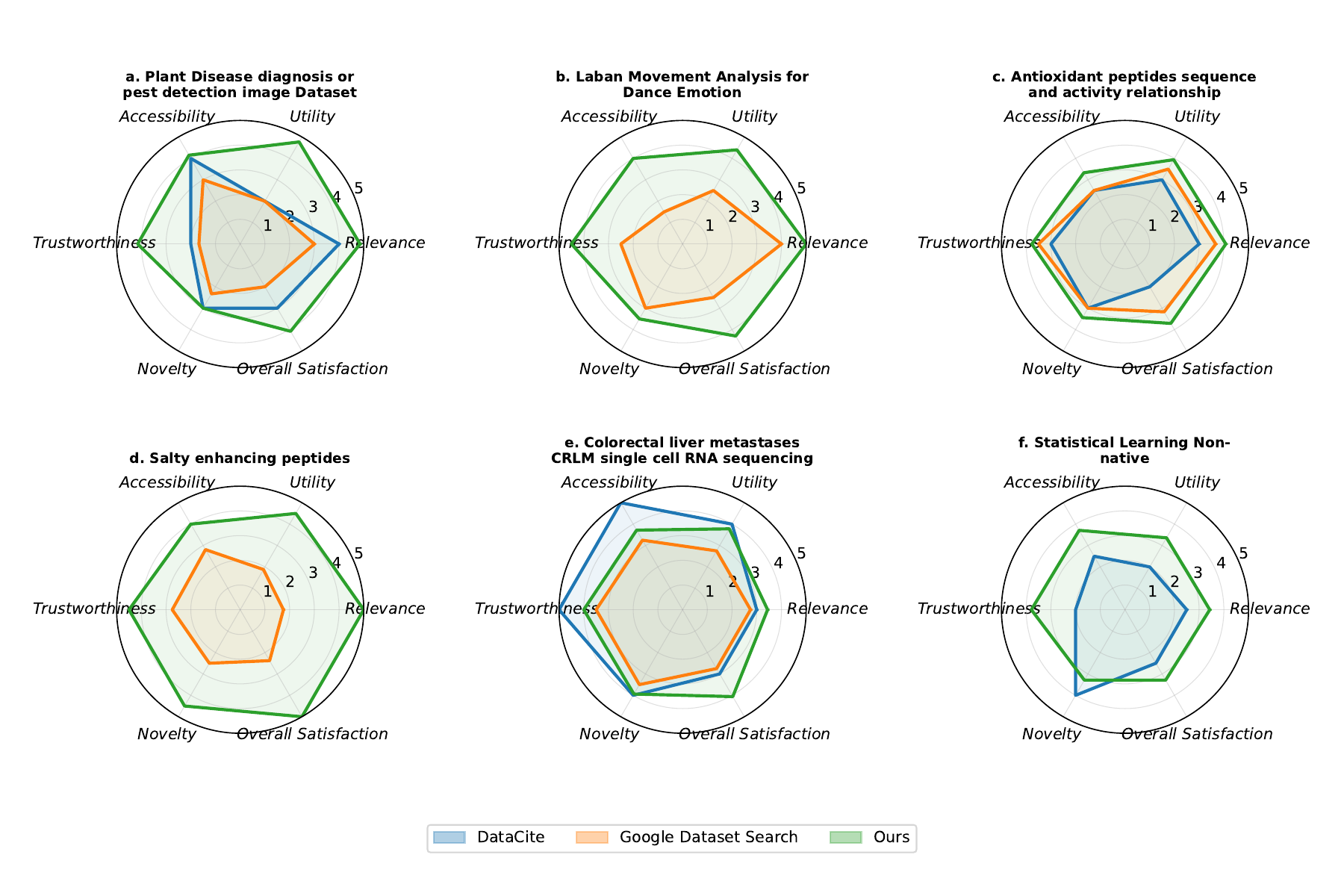}
\caption{Expert evaluation ratings across six cross-disciplinary research questions. Each radar chart visualizes the mean user ratings for a single query, comparing our system against the baselines across six dimensions of quality. The detailed numerical results are available in Table~\ref{tab:expert_all_rqs_single}.}
\label{fig:rq_radars}
\vspace{-5mm}
\end{figure*}

Experts rated datasets on six 5-point Likert dimensions: \textit{Relevance}, \textit{Utility}, \textit{Accessibility}, \textit{Trustworthiness}, \textit{Novelty}, and \textit{Overall Satisfaction}. 
Figure~\ref{fig:rq_radars} visualizes the per research question ratings, while Table~\ref{tab:expert_summary} presents aggregate results.
Per-Research Query mean~$\pm$~SD ratings are reported in the Appendix in Table~\ref{tab:expert_all_rqs_single}. 
Experts unanimously favored our system in \textit{Relevance} and \textit{Utility}, confirming that context-based retrieval identifies datasets genuinely useful for research. Gains were also consistent in \textit{Trustworthiness} and \textit{Overall Satisfaction}. In terms of \textit{Novelty}, experts indicated that a meaningful subset of the additional (non-survey) datasets surfaced by our system are new to them and still perceived as useful within their domains. 

Across all expert-rated candidates, our literature-driven method produced 105 datasets with complete \textit{Utility} and \textit{Novelty} scores, of which 45 (42.9\%) were rated at least~4 on both dimensions, compared to 4 of 31 (12.9\%) for Google Dataset Search and 2 of 6 (33.3\%) for DataCite Commons. 
These numbers quantitatively support that a substantial portion of the additional datasets surfaced by our approach are considered high utility by domain experts, and that a non-trivial subset is also regarded as novel for their specific research topics.

\begin{table}[!t]
\centering
\caption{Aggregate expert mean ratings (1--5).}
\label{tab:expert_summary}
\begin{tabular}{l|c|c|c}
\toprule
\textbf{Dimension} & \textbf{Ours} & \textbf{Google} & \textbf{DataCite} \\
\midrule
Relevance       & \textbf{4.33} & 3.07 & 2.60 \\
Utility         & \textbf{4.09} & 2.46 & 2.66 \\
Accessibility   & \textbf{3.80} & 2.87 & 3.17 \\
Trustworthiness & \textbf{4.13} & 2.49 & 2.84 \\
Novelty         & \textbf{3.64} & 2.92 & 3.50 \\
Overall         & \textbf{4.07} & 2.60 & 2.48 \\
\bottomrule
\end{tabular}
\end{table}

\subsubsection{Quality and Trustworthiness Analysis}

Beyond raw recall metrics, our evaluation reveals significant differences in the quality and trustworthiness of extracted datasets across systems.

\paragraph{Evidence Quality} our system achieves 14.23\% trusted-backed recall, significantly outperforming Google Dataset Search (1.63\%) and DataCite (0.00\%). This indicates that our neural extraction approach not only discovers more datasets but also identifies higher-quality resources with verifiable academic provenance.

\paragraph{Persistent Identifier Coverage} DataCite demonstrates the highest PID rate (87.50\%) due to its focus on formally published datasets with DOIs. However, this comes at the cost of extremely low recall (0.00\%), indicating that formal publication requirements severely limit dataset discovery scope. Our system balances PID coverage (68.52\%) with substantially higher recall, identifying both formally published and emerging datasets.

\paragraph{Redundancy and Efficiency} our dataset entity resolution pipeline achieves the lowest redundancy rate (0.26), demonstrating effective duplicate detection and consolidation. DataCite shows concerning redundancy (7.76), suggesting potential indexing inconsistencies, while Google maintains reasonable efficiency (0.29).

\subsubsection{System Scalability and Coverage}

Our approach demonstrates superior scalability, extracting 1,330 unique dataset entities compared to 79 for Google and 67 for DataCite. This 16.8× improvement in coverage reflects our system's ability to identify datasets embedded within citation contexts rather than relying solely on explicit metadata markup or formal registration.

\paragraph{Domain Adaptability} performance varies significantly across domains, with the highest recall in technical computer science areas (Multi-modal Knowledge Graph Reasoning: 81.82\%) and lower performance in interdisciplinary fields (Personalized Text Generation: 6.25\%). This reflects fundamental differences in research data practices: CS domains have established dataset sharing cultures, making citation-context mining effective, while interdisciplinary fields emphasize novel data collection over reuse. This validates our dual evaluation approach, as high recall may not correlate with practical research value across all domains.

\paragraph{FuzzyGain Robustness} our system exhibits positive FuzzyGain (1.71\%), indicating effective handling of dataset name variations through fuzzy matching. Baseline systems show zero FuzzyGain, suggesting limited robustness to naming inconsistencies common in scientific literature.

\subsubsection{Summary}
Across both automated and expert evaluations, grounding dataset discovery in \emph{citation context} yields resources judged more relevant, useful, and trustworthy than metadata-driven baselines. These findings validate our core hypothesis: literature-based semantic bridges between research questions and datasets enable more effective and generalizable dataset discovery across domains, expanding the actionable dataset space rather than merely increasing result volume.
The poor performance of baseline systems confirms that direct metadata mapping is insufficient for capturing the intent of complex research questions, necessitating the literature-driven approach proposed in this work.

\subsection{Case Studies}

\subsubsection{Dataset Family Resolution}
For research question: \textit{Document-level Event Extraction}.
Our system identified three ACE variants (\texttt{ACE}, \texttt{ACE 2005}, \texttt{ACE 2005 (zh)}) referencing the same catalog (\texttt{LDC2006T06}) but differing in scope and language. Traditional approaches either over-merge variants (losing semantic distinctions) or use paper DOIs as keys (causing entity misalignment).

Our solution employs family-level identifiers (catalog IDs, PIDs) as primary keys while preserving variant metadata as attributes. This enables accurate recall computation at the conceptual level while maintaining granular information, preventing recall inflation, and ensuring reproducible comparisons.

\section{Conclusion and Discussion}
This work introduced a literature-driven framework for dataset discovery that treats scientific publications as semantic bridges between research questions and data resources. The three-stage pipeline, comprising citation-context extraction, dataset recognition, and dataset-level entity resolution, demonstrates that grounding discovery in scholarly usage rather than metadata alone yields citation-verified datasets that are not only more relevant and credible but also richer in semantic context. Extracted directly from their textual and citation environments, these datasets convey signals about their purpose, methodology, and domain relevance, dimensions of meaning typically absent from metadata-based repositories.

A defining strength of the approach is that every discovered dataset has already undergone an implicit form of peer validation. By appearing in published research, each dataset has been described, evaluated, and applied within a scholarly workflow, reflecting its accessibility, usability, and reliability. This property renders the discovery process inherently quality-aware, reducing the burden of manual vetting. Moreover, by preserving the linguistic and conceptual context of dataset mentions, our framework exposes the reasoning behind dataset adoption, providing researchers with a more functional, context-rich understanding of the data landscape.

Empirical results demonstrate that the framework consistently outperforms metadata-dependent systems across diverse domains. These findings establish citation-context mining as a scalable and generalizable paradigm for discovery, capable of recovering both established datasets and previously undocumented yet valuable ones. By transforming static bibliographic references into actionable semantic links between research problems, data resources, and their documented applications, the framework advances a new mode of context-aware scientific search.

\paragraph{Limitations and Challenges}
while the framework achieves robust and meaningful discovery, several trade-offs naturally arise from its design. (1) Reliance on citation-rich literature favors datasets that have achieved visibility through prior publication and reuse. This ensures that identified datasets are credible and validated by the research community but may also limit coverage of newly released or niche datasets that have not yet been cited. Likewise, the temporal lag inherent in the publication cycle delays the inclusion of emerging datasets until they appear in subsequent works. (2) Dependence on large-scale open-access corpora such as S2AG introduces structural bias toward disciplines with established open-data practices, while underrepresenting fields dominated by paywalled publications. These are not flaws of the method but reflections of the broader data ecosystem. Future work integrating multi-source corpora and adaptive retrieval strategies can help mitigate these imbalances over time.

\paragraph{Ethical Considerations}
reliance on English-language literature and citation-based indexing may inadvertently underrepresent datasets originating from non-English research communities or from disciplines where data citation is less formalized. These biases arise not from the methodology itself but from the composition of the accessible scholarly record. Addressing them requires expanding coverage to multilingual and regional corpora and adapting extraction models to recognize diverse conventions of dataset acknowledgment across fields. Such extensions will promote a more balanced and globally representative discovery process.

\section*{Broader Impact and Future Directions}
The broader significance of this work lies in establishing citation-context mining as a generalizable paradigm for literature-grounded resource discovery. While our current focus is on datasets, the underlying principle naturally extends to other scholarly entities. 
For instance, the same methodology can be adapted to map the cross-disciplinary diffusion of computational software, trace the evolution of experimental protocols, and examine how theoretical frameworks are operationalized in empirical research. By capturing these diverse artifacts, future work can populate a comprehensive scholarly knowledge graph that links data, methods, and tools, thereby enhancing the understanding of the field. Integrating such a graph with large language models will enable natural-language querying of complex research relationships, thereby evolving the framework from a specific discovery tool into a unified model of scholarly knowledge interconnection that supports a transparent and reproducible scientific ecosystem.


\bibliographystyle{IEEEtran}
\bibliography{custom-slim}

\appendix
\section*{Detailed Pipeline Methodology}
This appendix details the end-to-end operational workflow of our literature-driven dataset discovery framework. The process is organized into three main phases.

\subsection{Stage 1: Scalable Corpus Preprocessing and Indexing}
\label{app:stage1}

This initial phase constructs a focused, query-relevant corpus of scientific text for analysis.

\paragraph{Seed Paper Retrieval}
given a natural-language RQ, we first query the Semantic Scholar API to retrieve a set of top K relevant seed papers (usually $200 \le K \le 400$). These papers form the initial, high-relevance core of our analysis corpus.

\paragraph{Citation Graph Expansion and Context Collection}
using our pre-indexed local snapshot of the S2AG corpus.
We collect all citation contexts where the seed papers are either citing another work or are being cited by another work. This expansion broadens the pool of relevant texts beyond the initial keyword match.

\paragraph{Offline Indexing}
to execute this phase at scale, our pipeline operates on a local S2AG snapshot. The raw data is converted to the Parquet columnar format and indexed using an embedded DuckDB
instance configured on a high-memory server (128 GB RAM, 32-thread parallelism, 1TB SSD Storage). We pre-materialize textual contexts for each citation link, allowing for near-instantaneous retrieval. This architecture reduced our median query latency from tens of seconds with a pure API approach to approximately 0.1 seconds, which is critical for interactive analysis.

\subsection{Stage 2: LLM-based Extraction and Relevance Filtering}
\label{app:stage2}
This phase uses an LLM in a multi-step process to identify dataset mentions and assess their relevance to the initial research question.

\begin{lstlisting}[mathescape, style=prompt, caption={Dataset Mention Extraction}, label={lst:prompt-extract}]
[System]
You are an expert scientific annotator. Extract dataset entities from the
given citation context. Rules:
- Output valid JSON only (no extra text).
- Ground findings strictly in the provided context; do not hallucinate.
- Controlled vocabularies:
  usage_role $\in$ {"Use","Modify","Evaluate Against"}
  content_type $\in$ {"Performed Work","Discovery","Produced Resource"}

[User]
Research Question (RQ):
"{RQ_TEXT}"
Citing Paper Title: "{CITING_TITLE}"
Cited Paper Title: "{CITED_TITLE}"
Citation Context (verbatim):
[BEGIN CONTEXT]
{CITATION_CONTEXT_TEXT}
[END CONTEXT]

Task:
1) Identify dataset/benchmark/corpus names explicitly or implicitly referenced.
2) For each dataset, provide:
   - name
   - usage_role
   - content_type
   - evidence (verbatim span)
   - confidence (0.0-1.0)
   - rationale (1-2 sentences grounded in context)
Output JSON only:
{
  "datasets": [
    {
      "name": "...",
      "usage_role": "Use|Modify|Evaluate Against",
      "content_type": "Performed Work|Discovery|Produced Resource",
      "evidence": "...",
      "confidence": 0.0-1.0,
      "rationale": "..."
    }
  ]
}
\end{lstlisting}

\paragraph{Dataset Mention Extraction}
each collected citation context is processed by our LLM (Qwen2.5-72B-Instruct) to identify mentions of datasets, benchmarks, or corpora. Using the SOFT classification framework, the prompt instructs the model to extract the dataset's name and its specific usage as described in the context.

\paragraph{Dataset Relevance Filtering}
a second LLM call is then made to filter the extracted datasets for relevance to the original research question. The model is provided with the RQ, the citation context, and the titles of both the citing and cited papers. It makes a binary decision (include/exclude) based on the semantic alignment between the dataset's usage and the research goal. This step filters out datasets that, while valid, are not pertinent to the user's query.
\begin{lstlisting}[style=prompt, caption={Relevance Filtering}, label={lst:prompt-rel}]
[System]
Decide whether a candidate dataset is relevant to the research question,
based solely on the provided context and titles. Respond with valid JSON only.

[User]
Research Question (RQ):
"{RQ_TEXT}"

Candidate Dataset:
"name": "{CANDIDATE_DATASET_NAME}"

Context:
[BEGIN CONTEXT]
{CITATION_CONTEXT_TEXT}
[END CONTEXT]

Citing Paper Title: "{CITING_TITLE}"
Citing Paper Abstract: "{CITING_ABSTRACT}"
Cited Paper Title: "{CITED_TITLE}"
Cited Paper Abstract: "{CITED_ABSTRACT}"

Decision rules:
- Relevant if the context shows the dataset was used, modified, or evaluated
  in pursuit of the RQ (or a directly aligned objective).
- Not relevant if usage is unrelated, purely background, or from a different domain.
- Be conservative; require explicit evidence in the context.
Output JSON only:
{
  "is_relevant": true/false,
  "confidence": 0.0-1.0,
  "reasoning": "brief, evidence-based justification"
}
\end{lstlisting}

\paragraph{Prompting and Infrastructure}
all LLM tasks run on a vLLM-powered inference server
with four NVIDIA A100-80GB GPUs. Our prompt engineering strategy casts the LLM as an expert annotator, providing strict inclusion/exclusion rules (e.g., extract \texttt{MIMIC-III}, ignore \texttt{BERT}) and mandating that all outputs are grounded in the provided text to prevent hallucination. A structured JSON schema (based on SOFT) is enforced for all outputs, and the model's temperature is set to 0.0 for deterministic results.

\subsection{Stage 3: Entity Resolution and Metadata Enrichment}
\label{app:stage3}

This phase cleans and consolidates the raw, filtered dataset mentions into a canonical list.

\paragraph{Normalization and Consolidation}
we apply a deterministic, rule-based procedure to resolve different surface forms (e.g., \texttt{ACE 2005 (zh)}, \texttt{ACE-2005}) to a single canonical entity. This involves strong lexical normalization (lowercasing, punctuation removal, etc.) followed by grouping mentions that share the same normalized form. Within each group, a human-readable display name is chosen by frequency, and all original surface forms are retained as aliases. All provenance is preserved to ensure the process is auditable.

\paragraph{Multi-Source URL and PID Retrieval}
we employ a three-tiered strategy to find a URL or PID for each dataset. First, we check if a URL was directly extracted from the citation context by the LLM. If not, we examine the metadata of the cited paper in S2AG; if it is a resource paper, we inherit its DOI. As a final step, we use the canonical dataset name to programmatically query an external source (Google Gemini search capabilities) to find its official homepage or repository.

\paragraph{Final Ranking and Output Generation}
the result of this entire pipeline is a clean, deduplicated, and ranked list of datasets relevant to the original research question. The final ranking is determined by the citation count: the number of unique papers within our retrieved corpus that were found to use a given dataset. This usage-based frequency serves as a proxy for the dataset's prevalence and importance in the context of the research question. Each entry in the final list includes the canonical dataset name, its usage context, the citation count, and a verifiable URL or PID.

\section*{Per-Query Expert Ratings}
\label{app:expert_details}

\begin{table}[H]
\centering
\caption{Consolidated expert evaluation ratings across all cross-disciplinary research questions. Each block shows the mean user ratings (1--5) $\pm$ standard deviation for a query.}
\label{tab:expert_all_rqs_single}
\setlength{\tabcolsep}{3.5pt} 
\resizebox{\columnwidth}{!}{%
\begin{tabular}{@{}lcccccc@{}}
\toprule
\textbf{System} & \textbf{Relevant} & \textbf{Utility} & \textbf{Accessible} & \textbf{Trustworthy} & \textbf{Novelty} & \textbf{Overall} \\
\midrule

\multicolumn{7}{l}{\textbf{\makecell[l]{Agricultural Sciences: \\ Plant Disease diagnosis or pest detection image Dataset}}} \\ \addlinespace
Ours & 4.38 $\pm$ 0.52 & 4.00 $\pm$ 0.63 & 3.83 $\pm$ 0.75 & 4.17 $\pm$ 0.75 & 3.67 $\pm$ 0.82 & 4.00 $\pm$ 0.63 \\
Google & 2.88 $\pm$ 0.64 & 2.56 $\pm$ 0.73 & 2.88 $\pm$ 0.64 & 2.56 $\pm$ 0.73 & 2.88 $\pm$ 0.64 & 2.56 $\pm$ 0.73 \\
DataCite & 2.63 $\pm$ 0.74 & 2.69 $\pm$ 0.70 & 3.19 $\pm$ 0.54 & 2.81 $\pm$ 0.66 & 3.50 $\pm$ 0.60 & 2.56 $\pm$ 0.66 \\
\midrule

\multicolumn{7}{l}{\textbf{Humanities (Arts): Laban Movement Analysis for Dance Emotion}} \\ \addlinespace
Ours & 4.67 $\pm$ 0.52 & 4.17 $\pm$ 0.75 & 3.67 $\pm$ 0.82 & 4.17 $\pm$ 0.75 & 3.50 $\pm$ 0.84 & 4.00 $\pm$ 0.89 \\
Google & 3.67 $\pm$ 0.82 & 2.67 $\pm$ 0.82 & 2.83 $\pm$ 0.75 & 2.50 $\pm$ 0.84 & 3.00 $\pm$ 0.89 & 2.50 $\pm$ 0.84 \\
DataCite & 2.50 $\pm$ 0.84 & 2.67 $\pm$ 0.52 & 3.17 $\pm$ 0.41 & 2.67 $\pm$ 0.52 & 3.50 $\pm$ 0.55 & 2.50 $\pm$ 0.55 \\
\midrule

\multicolumn{7}{l}{\textbf{\makecell[l]{Engineering and Technology (Food): \\ Antioxidant peptides sequence and activity relationship}}} \\ \addlinespace
Ours & 4.75 $\pm$ 0.35 & 4.43 $\pm$ 0.53 & 3.94 $\pm$ 0.70 & 4.31 $\pm$ 0.60 & 3.63 $\pm$ 0.80 & 4.38 $\pm$ 0.56 \\
Google & 2.88 $\pm$ 0.64 & 2.63 $\pm$ 0.74 & 2.88 $\pm$ 0.64 & 2.69 $\pm$ 0.70 & 2.88 $\pm$ 0.64 & 2.69 $\pm$ 0.70 \\
DataCite & 2.44 $\pm$ 0.73 & 2.69 $\pm$ 0.70 & 3.13 $\pm$ 0.64 & 2.75 $\pm$ 0.71 & 3.50 $\pm$ 0.76 & 2.69 $\pm$ 0.74 \\
\midrule

\multicolumn{7}{l}{\textbf{Engineering and Technology (Food): Salty Enhancing Peptides}} \\ \addlinespace
Ours & 4.89 $\pm$ 0.20 & 4.78 $\pm$ 0.33 & 4.40 $\pm$ 0.52 & 4.67 $\pm$ 0.33 & 3.90 $\pm$ 0.57 & 4.78 $\pm$ 0.33 \\
Google & 2.10 $\pm$ 0.57 & 2.20 $\pm$ 0.42 & 2.50 $\pm$ 0.53 & 2.30 $\pm$ 0.48 & 2.50 $\pm$ 0.53 & 2.20 $\pm$ 0.42 \\
DataCite & 2.00 $\pm$ 0.47 & 2.30 $\pm$ 0.48 & 2.40 $\pm$ 0.52 & 2.50 $\pm$ 0.53 & 4.10 $\pm$ 0.32 & 2.30 $\pm$ 0.48 \\
\midrule

\multicolumn{7}{l}{\textbf{\makecell[l]{Medical and Health Sciences (Clinical): \\ Colorectal liver metastases CRLM single cell RNA sequencing}}} \\ \addlinespace
Ours & 4.33 $\pm$ 0.58 & 4.00 $\pm$ 0.63 & 3.83 $\pm$ 0.75 & 4.17 $\pm$ 0.75 & 3.67 $\pm$ 0.82 & 4.00 $\pm$ 0.63 \\
Google & 3.00 $\pm$ 0.63 & 2.67 $\pm$ 0.52 & 2.83 $\pm$ 0.75 & 2.50 $\pm$ 0.55 & 3.00 $\pm$ 0.63 & 2.50 $\pm$ 0.55 \\
DataCite & 2.67 $\pm$ 0.52 & 2.67 $\pm$ 0.52 & 3.17 $\pm$ 0.41 & 2.83 $\pm$ 0.41 & 3.50 $\pm$ 0.55 & 2.50 $\pm$ 0.55 \\
\midrule

\multicolumn{7}{l}{\textbf{Social Sciences (Educational): Statistical Learning Non-native}} \\ \addlinespace
Ours & 4.25 $\pm$ 0.96 & 3.75 $\pm$ 0.96 & 3.75 $\pm$ 0.96 & 4.00 $\pm$ 0.82 & 3.75 $\pm$ 0.96 & 3.75 $\pm$ 0.96 \\
Google & 3.00 $\pm$ 0.82 & 2.75 $\pm$ 0.96 & 2.75 $\pm$ 0.96 & 2.75 $\pm$ 0.96 & 3.00 $\pm$ 0.82 & 2.75 $\pm$ 0.96 \\
DataCite & 2.75 $\pm$ 0.96 & 2.75 $\pm$ 0.96 & 3.25 $\pm$ 0.96 & 3.00 $\pm$ 0.82 & 3.50 $\pm$ 0.58 & 2.75 $\pm$ 0.96 \\
\bottomrule
\end{tabular}%
}
\end{table}

\end{document}